\documentclass[english,10pt,prl,showkeys]{revtex4}
\usepackage[latin9]{inputenc}
\usepackage{amsmath}

\makeatletter
\usepackage{babel}
\makeatother

\begin{document}

\title{Three Views of a Secret in Relativistic Thermodynamics}

\author{Tadas K. Nakamura}

\affiliation{CFAAS, Fukui Prefectural University}

\email{tadas@fpu.ac.jp}


\begin{abstract}
It has been shown three different views in relativistic thermodynamics
can be derived from the basic formulation proposed by van Kampen and
Israel. The way to decompose energy-momentum into the reversible and
irreversible parts is not uniquely determined, and different choices
result in different views. The effect of difference in the definition
of a finite volume is also considered.
\end{abstract}
\maketitle
There has been long controversy about the relativistic thermodynamics.
A number of theories were proposed in 1960s, and the discussion seems
to have arrived at a vague general agreement that each theory is consistent
in its own framework by early 1970s \citet{yuen70}. However, papers
has been still published long after that, even until today (e.g.,
\citet{aresdeparga05,requardt08}), proposing new formulations which
are allegedly better than others.

Roughly speaking there are three different views on the relativistic
thermodynamics, which are characterized by the difference in the temperature
$T$ of a moving body in the following:\begin{equation}
\begin{cases}
\textnormal{I)} & T=T_{0}\gamma^{-1}\,,\\
\textnormal{II)} & T=T_{0}\gamma\,,\\
\textnormal{III)} & T=T_{0}\,.\end{cases}\label{eq:views}\end{equation}
where $T_{0}$ is the temperature measured in a frame comoving with
the body, $\gamma$ is the Lorentz factor defined as $\gamma=1/\sqrt{1-v^{2}}$
with the speed of the body $v$ relative to the rest frame (we use
the unit of $c=1$). 

Most of theories proposed so far can be categorized into one of the
above three. Papers published right after the establishment of special
relativity (e.g., \citet{einstein07,planck08}) are based on View
I. Theories with View II were extensively investigated in the middle
of 1960s (e.g, \citet{gamba65,kibble66}) stimulated by the papers
by \citet{ott63} and \citet{arzelies65}. A little later a theory
in View III was proposed by  \citet{lansberg66}. 

There is another theory by \citet{vankampen68} in View III; his stand
point is quite different from other theories in Views I, II, and III.
He treats the three components of the velocity as thermodynamical
parameters in addition to the temperature. The van Kampen's theory
was later refined by \citet{israel76} into a more transparent form. 

The author of the present paper strongly believes this van Kampen-Israel
theory is the one in \emph{the book}, i.e., the very fundamental one
based on which other formulations can be derived. The purpose of the
present paper is to show the above three views can be actually derived
from the van Kampen-Israel theory. The great advantage of the van
Kampen-Israel theory is in the point that it does not need the concept
of heat or work. The second law can be expressed with well defined
mechanical quantities such as the energy-momentum or four velocity.
\bigskip{}

The difference of the above three views mainly comes from the difference
in the definition of heat. Non-relativistic thermodynamics decomposes
the energy increase $\Delta E$ into two parts, heat $\Delta Q$ and
work $\Delta W$ namely, as $\Delta E=\Delta Q+\Delta W$. Since the
energy is one component of the energy-momentum four vector, the decomposition
should be expressed in the form of four vectors in the relativistic
thermodynamics:\begin{equation}
\Delta\bar{G}=\Delta\bar{Q}+\Delta\bar{W}\,,\label{eq:heatwork}\end{equation}
where $\Delta\bar{G}$ is the change of the energy-momentum, and its
reversible and irreversible parts are expressed as $\Delta\bar{W}$
and $\Delta\bar{Q}$; we denote a four vector as a whole by a bar
(e.g., $\bar{G}$) and its each component by indices (e.g., $G^{\mu}$)
in this paper.

Most of theories in Views I and II determine the temperature from
the following entropy expression. \begin{equation}
\Delta S=\frac{\Delta Q}{T}\,,\label{eq:entropy}\end{equation}
where $\Delta Q$ is the temporal component of $\Delta\bar{Q}$. However,
there is ambiguity in the decomposition in  (\ref{eq:heatwork}) and
the heat is not uniquely determined as we will see in the present
paper. Views I and II define the heat as $\Delta Q=\Delta Q_{0}\gamma^{-1}$
and $\Delta Q=\Delta Q_{0}\gamma$ ($\Delta Q_{0}$ is the heat measured
in the comoving frame) respectively, and both definitions are consistent
as the temporal component of an irreversible energy-momentum change.
Consequently two different temperatures (Views I and II) are derived
from  (\ref{eq:views}) since the entropy is supposed to be Lorentz
invariant. View III tries to accommodate both somehow.

The author considers this difference of $\Delta\bar{Q}$ is the main
reason for the confusion in relativistic thermodynamics. However,
the definition of a finite volume may also make the problem complicated.
This point has been known since the very early years of relativity
(\citet{fermi23}), and must have been well recognized during the
controversy in 1960s (\citet{gamba65,kibble66,yuen70}). However,
curiously enough, its importance is not well understood and a number
of erroneous statements on this point are found in papers since 1960s
to this date. 

We will see in the present paper these confusions can be cleared by
the covariant expression of the van Kampen-Israel theory. For this
purpose it is convenient to define the volume as a four vector (\citet{tadas06}):\begin{equation}
V^{\eta}(\bar{w})=\frac{w^{\eta}V_{0}}{w^{\mu}u_{\mu}}\,.\label{eq:volume}\end{equation}
This four vector represents a space-like volume orthogonal to the
unit vector $w^{\eta}$; in other words, this vector defines the volume
viewed in a reference frame with the four velocity $\bar{w}$. \bigskip{}

The van Kampen-Israel theory defines the entropy change of a matter
with a finite volume as\begin{equation}
\Delta S=\beta_{0}u_{\mu}V^{\nu}\Delta T_{\nu}^{\mu}-\beta_{0}u_{\mu}P\Delta V^{\mu}\,,\label{eq:no5}\end{equation}
where $P$, $T_{\nu}^{\mu}$ are the pressure and energy momentum
tensor respectively, and $\beta_{0}=1/T_{0}$ is the inverse temperature
measured in the rest frame.

When we define the total energy-momentum four vector as $G^{\mu}(\bar{w})=V^{\nu}(\bar{w})\, T_{\nu}^{\mu}$
then  (\ref{eq:no5}) can be expressed as\begin{equation}
\Delta S=\beta_{0}u_{\mu}[\Delta G^{\mu}(\bar{w})-P\Delta V^{\mu}(\bar{w})]\,.\end{equation}
Both $\Delta\bar{G}$ and $\Delta\bar{V}$ depends on $\bar{w}$,
i.e., the direction of the volume in the Minkowski space, however,
the dependence is canceled out by taking inner product with $u_{\mu}$
and $\Delta S$ becomes invariant.

The heat/work is defined as an irreversible/reversible part of the
energy-momentum $\Delta\bar{G}$ in  (\ref{eq:heatwork}), which means
\begin{equation}
\beta_{0}u_{\mu}\Delta Q^{\mu}>0\,,\,\,\,\beta_{0}u_{\mu}[\Delta W^{\mu}-P\Delta V^{\mu}]=0\,.\label{eq:irreversible}\end{equation}
Obviously the above conditions cannot determine the heat and work
uniquely; when we define new values of the heat and work by $\Delta\bar{Q}'=\Delta\bar{Q}+\bar{A}$
and $\Delta\bar{W}'=\Delta\bar{W}-\bar{A}$ with an arbitrary four
vector $\bar{A}$ that satisfies $u_{\mu}A^{\mu}=0$,  (\ref{eq:irreversible})
holds for the new values $\Delta\bar{Q}'$ and $\Delta\bar{W}'$.
This ambiguity causes the difference in  (\ref{eq:views}) as we will
see in the following.\bigskip{}

Suppose a matter moving in the $x$ direction with a four velocity
$(u_{t},u_{x},0,0)$. Then energy momentum tensor may be written in
the rest frame as\[
T_{\mu\nu}=\left(\begin{array}{cc}
u_{t}^{2}\varepsilon_{0}+u_{x}^{2}P & u_{t}u_{x}(\varepsilon_{0}+P)\\
u_{t}u_{x}(\varepsilon_{0}+P) & u_{x}^{2}\varepsilon_{0}+u_{t}^{2}P\end{array}\right)\,,\]
with $\varepsilon_{0}$ being the energy density measured in the comoving
frame. We ignore the dimension in $y$ and $z$ direction for simplicity.
Note that $\varepsilon_{0}$ and $P$ do not depend on $t$ or $x$
because the matter is in the equilibrium state. 

We introduce a parameter $\theta=\tanh^{-1}(w^{x}/w^{t})$ to define
the volume in  (\ref{eq:volume}). Then the total energy-momentum
can be expressed as a function of $\theta$ in the following:\begin{equation}
\bar{G}(\theta)=\left(\begin{array}{c}
E(\theta)\\
G(\theta)\end{array}\right)=\left(\begin{array}{c}
E_{0}\cosh\alpha+{\displaystyle PV_{0}\sinh\alpha\tanh(\theta-\alpha)}\\
E_{0}\sinh\alpha+{\displaystyle PV_{0}\cosh\alpha\tanh(\theta-\alpha)}\end{array}\right)\label{eq:energymomentum}\end{equation}
where $E_{0}=\varepsilon_{0}V_{0}$ is the total energy measured in
the comoving frame, and the velocity of the matter is parametrized
by $\alpha=\tanh^{-1}(u^{x}/u^{t})$ instead of $\bar{u}$.

We need another parameter to fix the ambiguity of heat in  (\ref{eq:heatwork}).
Let us introduce a parameter $\phi$ such that \[
\Delta\bar{Q}=\left(\begin{array}{c}
\Delta Q\\
\Delta Q\tanh\phi\end{array}\right)\]
to this end. This parameter $\phi$ specifies the frame in which the
heat is purely timelike, in other words, the frame in which the heat
looks {}``heat'' only without momentum. The rest frame and comoving
frame are represented by $\phi=0$ and $\phi=\alpha$ respectively.

The work $\Delta\bar{W}$ is then calculated as 

\[
\Delta\bar{W}=\left(\begin{array}{c}
\Delta E_{0}\cosh\alpha+{\displaystyle \Delta(PV_{0})\sinh\alpha\tanh(\theta-\alpha)-\Delta Q}\\
\Delta E_{0}\sinh\alpha+{\displaystyle \Delta(PV_{0})\cosh\alpha\tanh(\theta-\alpha)}-\Delta Q\tanh\phi\end{array}\right)\,.\]
Since the work $\Delta\bar{W}$ must satisfy  (\ref{eq:irreversible}),
the heat $\Delta Q$ is uniquely determined when $\phi$ is given:\begin{equation}
\Delta Q(\phi)=\frac{\cosh\phi}{\cosh(\phi-\alpha)}\,\Delta Q_{0}\,,\label{eq:heat}\end{equation}
where $\Delta Q_{0}=\Delta E_{0}-P\Delta V_{0}$. 

\bigskip{}

Various formulations can be derived by expressing $\Delta Q(\phi)$
in  (\ref{eq:heat}) with the energy-momentum $\bar{G}(\theta)$ in
 (\ref{eq:energymomentum}) by choosing different $\phi$ and $\theta$.
Any value of $\phi$ and $\theta$ can determine the relativistic
thermodynamical equation in general, however, the value of the rest
frame or the comoving frame ($0$ or $\alpha$) are practically preferable
choices. In the following we examine three typical choices in Views
I, II, and III.

Typical theories choose the same value for $\phi$ and $\theta$ ($\phi=\theta$)
because they consider the heat exchange and volume change in the same
frame %
\footnote{This is typical, but not always. For example, the result by \citet{kibble66}
can be derived with $\phi=\alpha$ and $\theta=0$.%
}. For example, \citet{ott63} assumes a Carnot cycle in the comoving
frame; the steps in the cycle, including the heat exchange and volume
change, take place in the moving frame. Then  (\ref{eq:heat}) can
be cast in the form of\[
Q(\theta)=\frac{\cosh\theta}{\cosh(\theta-\alpha)}\Delta E_{0}-P\Delta V^{t}(\theta)\,.\]
The temporal component $\Delta V^{t}(\theta)$ is regarded as the
volume change in these theories, and denoted simply by $\Delta V$
and little attention has been paid to its dependence on $\theta$.
Then the above equation can be regarded as to correspond to the definition
of heat $\Delta Q=\Delta E-P\Delta V$ in non-relativistic thermodynamics.
The coefficient of the $\Delta E_{0}$ term determines the transformation
rule of the heat, and consequently, that of the temperature.

View I is typically derived with $\phi=\theta=0$, which means both
the heat and volume are defined in the rest frame. This choice gives
$\Delta Q=\Delta Q_{0}\gamma^{-1}$ and thus $T=T_{0}\gamma^{-1}$
because of (\ref{eq:entropy}). The calculation of heat is obtained
by subtracting $\gamma^{-1}u_{x}\Delta G$ from the energy, regarding
this as the work to cause acceleration:\begin{equation}
\Delta Q(0)=E(0)-\gamma^{-1}u_{x}\Delta G(0)-P\Delta V(0)\,,\label{eq:Q1}\end{equation}
where we write $\Delta V=\Delta V^{t}$.

The typical choice of View II is $\phi=\theta=\alpha$, resulting
$T=T_{0}\gamma$. The heat and the volume are defined in the comoving
frame, and the expression in the rest frame is a result of their Lorentz
transform. The calculation of $\Delta Q$ is straightforward:\begin{equation}
\Delta Q(\alpha)=\Delta E(\alpha)-P\Delta V(\alpha)\,.\label{eq:Q2}\end{equation}

\citet{lansberg66} considered the temperature must be a Lorentz-invariant
as in View III from the symmetry. When two identical systems are moving
relative to each other, there is no reason for one system to have
a temperature higher than the others'. This argument can be represented
by choosing $\phi=\frac{1}{2}\alpha$ in  (\ref{eq:heat}) to treat
the rest frame and the comoving frame symmetrically. It should be
noted that his actual calculation is more complicated, but we do not
examine its details here.

\bigskip{}
In the present paper we have successfully derived three different
views of relativistic thermodynamics from one basic formulation proposed
by \citet{vankampen68} and \citet{israel76}. The difference comes
from two factors, the definitions of the heart and volume namely,
which are represented by the two parameters $\phi$ and $\theta$
here. The papers published so far on this topic are so numerous that
it is not practical to check all of them. However, the author believes
all the formulations can be derived from the van Kampen-Israel theory
as long as they are not wrong.

\bibliographystyle{apsrev}

\bibliography{tvs}

\begin{thebibliography}{14}
\expandafter\ifx\csname natexlab\endcsname\relax\def\natexlab#1{#1}\fi
\expandafter\ifx\csname bibnamefont\endcsname\relax
  \def\bibnamefont#1{#1}\fi
\expandafter\ifx\csname bibfnamefont\endcsname\relax
  \def\bibfnamefont#1{#1}\fi
\expandafter\ifx\csname citenamefont\endcsname\relax
  \def\citenamefont#1{#1}\fi
\expandafter\ifx\csname url\endcsname\relax
  \def\url#1{\texttt{#1}}\fi
\expandafter\ifx\csname urlprefix\endcsname\relax\def\urlprefix{URL }\fi
\providecommand{\bibinfo}[2]{#2}
\providecommand{\eprint}[2][]{\url{#2}}

\bibitem[{\citenamefont{Yuen}(1970)}]{yuen70}
\bibinfo{author}{\bibfnamefont{C.~K.} \bibnamefont{Yuen}},
  \bibinfo{journal}{Amer.\ J.\ Phys.} \textbf{\bibinfo{volume}{38}},
  \bibinfo{pages}{246} (\bibinfo{year}{1970}).

\bibitem[{\citenamefont{Requardt}()}]{requardt08}
\bibinfo{author}{\bibfnamefont{M.}~\bibnamefont{Requardt}},
  \eprint{arXiv:0801.2639}.

\bibitem[{\citenamefont{de~Parga et~al.}(2005)\citenamefont{de~Parga,
  Lop{\'e}z-Carrera, and Anulo-Brown}}]{aresdeparga05}
\bibinfo{author}{\bibfnamefont{G.~A.} \bibnamefont{de~Parga}},
  \bibinfo{author}{\bibnamefont{Lop{\'e}z-Carrera}}, \bibnamefont{and}
  \bibinfo{author}{\bibfnamefont{F.}~\bibnamefont{Anulo-Brown}},
  \bibinfo{journal}{J.\ Math.\ Phys.} \textbf{\bibinfo{volume}{38}},
  \bibinfo{pages}{2821} (\bibinfo{year}{2005}).

\bibitem[{\citenamefont{Einstein}(1907)}]{einstein07}
\bibinfo{author}{\bibfnamefont{A.}~\bibnamefont{Einstein}},
  \bibinfo{journal}{Jb.\ Radioaktivitat} \textbf{\bibinfo{volume}{4}},
  \bibinfo{pages}{411} (\bibinfo{year}{1907}).

\bibitem[{\citenamefont{Planck}(1908)}]{planck08}
\bibinfo{author}{\bibfnamefont{M.}~\bibnamefont{Planck}},
  \bibinfo{journal}{Ann.\ d.\ Phys.} \textbf{\bibinfo{volume}{76}},
  \bibinfo{pages}{1} (\bibinfo{year}{1908}).

\bibitem[{\citenamefont{Gamba}(1965)}]{gamba65}
\bibinfo{author}{\bibfnamefont{A.}~\bibnamefont{Gamba}},
  \bibinfo{journal}{Nuovo Cimento} \textbf{\bibinfo{volume}{37}},
  \bibinfo{pages}{1792} (\bibinfo{year}{1965}).

\bibitem[{\citenamefont{Kibble}(1966)}]{kibble66}
\bibinfo{author}{\bibfnamefont{T.~W.~B.} \bibnamefont{Kibble}},
  \bibinfo{journal}{Nuovo Cimento} \textbf{\bibinfo{volume}{41B}},
  \bibinfo{pages}{167} (\bibinfo{year}{1966}).

\bibitem[{\citenamefont{Ott}(1963)}]{ott63}
\bibinfo{author}{\bibfnamefont{H.}~\bibnamefont{Ott}}, \bibinfo{journal}{Z.\
  Physik} \textbf{\bibinfo{volume}{175}}, \bibinfo{pages}{70}
  (\bibinfo{year}{1963}).

\bibitem[{\citenamefont{Arzelies}(1965)}]{arzelies65}
\bibinfo{author}{\bibfnamefont{H.}~\bibnamefont{Arzelies}},
  \bibinfo{journal}{Nuovo Cimento} \textbf{\bibinfo{volume}{35}},
  \bibinfo{pages}{792} (\bibinfo{year}{1965}).

\bibitem[{\citenamefont{Lansberg}(1966)}]{lansberg66}
\bibinfo{author}{\bibfnamefont{P.~T.} \bibnamefont{Lansberg}},
  \bibinfo{journal}{Nature} \textbf{\bibinfo{volume}{212}},
  \bibinfo{pages}{571} (\bibinfo{year}{1966}).

\bibitem[{\citenamefont{van Kampen}(1968)}]{vankampen68}
\bibinfo{author}{\bibfnamefont{N.~G.} \bibnamefont{van Kampen}},
  \bibinfo{journal}{Phys.\ Rev.} \textbf{\bibinfo{volume}{173}},
  \bibinfo{pages}{295} (\bibinfo{year}{1968}).

\bibitem[{\citenamefont{Israel}(1976)}]{israel76}
\bibinfo{author}{\bibfnamefont{W.}~\bibnamefont{Israel}},
  \bibinfo{journal}{Ann.\ Phys.} \textbf{\bibinfo{volume}{106}},
  \bibinfo{pages}{310} (\bibinfo{year}{1976}).

\bibitem[{\citenamefont{Fermi}(1923)}]{fermi23}
\bibinfo{author}{\bibfnamefont{E.}~\bibnamefont{Fermi}},
  \bibinfo{journal}{Nuovo Cimenmto} \textbf{\bibinfo{volume}{25}},
  \bibinfo{pages}{159} (\bibinfo{year}{1923}).

\bibitem[{\citenamefont{Nakamura}(2006)}]{tadas06}
\bibinfo{author}{\bibfnamefont{T.~K.} \bibnamefont{Nakamura}},
  \bibinfo{journal}{Phys.\ Lett.\ A} \textbf{\bibinfo{volume}{352}},
  \bibinfo{pages}{175} (\bibinfo{year}{2006}), \eprint{arXiv:physics/0505004}.

\end{thebibliography}

\end{document}